\documentclass[12pt]{article}
\usepackage[left=2cm,right=2cm,top=2cm,bottom=2cm,bindingoffset=0cm]{geometry}
\usepackage{epsfig}
\usepackage{epstopdf}
\usepackage{multirow}
\renewcommand{\Ref}[1]{(\ref{#1})}
\newcommand{\beao}{\begin{eqnarray*}}
\newcommand{\eeao}{\end{eqnarray*}}
\newcommand{\be}{\begin{equation}}
\newcommand{\ee}{\end{equation}}
\newcommand{\bea}{\begin{eqnarray}}
\newcommand{\eea}{\end{eqnarray}}
\newcommand{\beq}{\begin{eqnarray}}
\newcommand{\eeq}{\end{eqnarray}}
\newcommand{\nn}{\nonumber}

\begin{document}

\title{$A_0$ condensation,  Nielsen's identity \\and  effective potential of order parameter}
\author{
 V. Skalozub \thanks{e-mail: Skalozubv@daad-alumni.de}\\
{\small Oles Honchar Dnipro National University, 49010 Dnipro, Ukraine}}

\date{\small April 7, 2021}

\maketitle.

\begin{abstract}
 In  high temperature SU(2) gluodynamics,    the condensation of the zero component gauge field potential $ A_0 = const$  and its gauge-fixing dependence   are investigated. The $ A_0$  is mutually related with Polyakov's loop $  \left <  L \right> .$    The two-loop effective potential $W(A_0, \xi)$   is recalculated in the background  relatvistic $R_\xi$ gauge.     It depends on the parameter $\xi$, has a nontrivial minimum and  satisfies Nielsen's identity. These  signs mean gauge invariance of the   condensation phenomenon.

 We express $ W(A_0, \xi)$  in terms of   $  \left <  L \right> $   and obtain  the effective potential of order parameter $W(A_0^{cl})$ which is  independent of   $\xi$ and  has a nontrivial minimum position. Hence the $A_0^{cl} $ condensate value is detected.  We show that the equation  relating $ A_0$ and observable $A_0^{cl}$ coincides with the special  characteristic orbit in the $(A_0, \xi$)-plain along which the $W(A_0, \xi)$ is   $\xi$-independent.    In this way the link between these two gauge invariant  descriptions is established.  The minimum value of   Polyakov's loop,    the two-loop Debye mass and thermodynamical pressure are calculated. We also shown that  $A_0^{cl}$ condensate stabilizes the charged gluon spectrum in the chromomagneric field $H(T)$ which is spontaneously generated at high temperature due to Savvidy's mechanism. These two fields form an actual background of plasma.  Comparison with  results of other authors  (and other approaches) is given.

Key words: $A_0$ condensate; Nielsen's identity; effective potential; order parameter; gauge invariance, spontaneous magnetization.
\end{abstract}
\section{Introduction}
Investigations of the deconfinement phase transition (DPT) and)
quark-gluon plasma (QGP)   are in the center of modern high energy
physics. This matter state creates  at high temperature due to
asymptotic freedom of strong interactions. In
quantum field theory,  the order parameter for the DPT is
Polyakov's loop (PL), which is zero at low temperature and nonzero
at high temperatures $T > T_d $, where $T_d \sim 160-180 $ MeV
 is the DPT temperature (for critical temperature estimations see, for instance, \cite{aoki09-6-088}). Standard
information on the DPT is adduced, in particular, in
\cite{gros81-53-43}, \cite{kapu89b}, \cite{bellac96b}.

    In $SU(2)$,  the PL is defined as (in general case see \cite{sred81-179-239}):
\be \label{PL} \left <  L \right>  = \frac{1}{2}  Tr P \exp\bigl ( i g \int_0^\beta d x_4 ~A_0(x_4, \vec{x})\bigr) . \ee
Here, $g$ is coupling constant,  $A_0(x_4, \vec{x})  = A^a_0(x_4, \vec{x} ) \frac{\tau^a}{2}$ is the zero component   of the gauge field
potential, $\tau^a$ is Pauli's matrix,  $\beta$ is inverse temperature and  integration  is going along the fourth direction in  the  Euclidean space-time.

 The PL  violates the  color center group symmetry $Z(2)$ that could
result in spontaneous generation of color static  potential $A_0^3 = const$  and   non conservation of  color charge $Q^3$.
 Investigations of these phenomena were started for forty years ago and have been carried out by   different approaches. Among them calculation  of the  effective potentials in quantum field theory in continuum and on the lattice, Monte Carlo lattice simulations.
 The $A_0$ condensate is important dynamical parameter regulating infrared region of momenta of gauge fields at finite temperature. First it has been derived within two-loop effective potential (EP) $W(A_0, T)$ \cite{anis84-10-439}. Practically simultaneously it was observed that in a relativistic background $R_\xi$ gauge the EP and the  value of the condensed field detected as its minimum position $(A_0)_{min}$ are gauge-fixing dependent. That  called numerous discussions about gauge invariance of the $A_0$ condensation phenomenon as whole. In Refs. \cite{bely91-254-153}, \cite{enqv90-47-291} it was claimed that only zero condensate value is compartible with $\xi$-independence.   The other conclusion  was found by the present author within  Nielsen's ilentity approach \cite{skal92-7-2895}, \cite{skal94-50-1150}. It has been demonstrated that $W (A_0, \xi)$   satisfies Nielsen's identity  \cite{niel75-101-173}, \cite{fuku76-13-3469}.    Hence gauge invariance follows. The review paper on these and related calculations  and obtained results is \cite{bori95-43-301}.

  Recently in Ref. \cite{kort20-803-135336} for  $SU(N)$  gluodynamics  it has been found a nonperturbative  procedure for  removing  $\xi$-dependency within a constrained  potential method. It includes special resummations of perturbation series.  The obtained effective potential  is gauge-fixing independent and has a nontrivial minimum, that proves  gauge invariance of the condensate. In Ref.\cite{rein15-742-61} the DPT was investigated  in perturbation theory  in $SU(2)$ gluodynamics in a modified Landau-DeWitt gauge and the phase transition type, in particular, was determined. In Ref.\cite{rein16-93-105002} it was generalized to $SU(3)$ Yang-Mills theory.

 The  above mentioned results, calculated by different methods, and existing in the literature discrepancies stimulate an investigation, which  explains the  latter and  relates  the former ones. This is the main goal of the present paper. To realize that we derive  a link  between  Nielsen's identity for the two-loop EP \cite{skal92-7-2895}  and Belyaev's   EP of order parameter \cite{bely91-254-153}. This is  key point because, as we  show below,  a very perspective idea to express the EP $W(A_0, \xi)$ in terns of the observable gauge invariant parameter   $\left <  L \right> $ was  realized  with calculation errors. Hence incorrect conclusion about the absence of the condensation at two-loop order  followed.

Since necessary for what follows standard calculations have been reported in numerous papers, we adduce  them in brief and concentrate on the most important points. In next section, to have  consistent presentation, we give information on the EP and Nielsen's identity calculations. In sect. 3 we calculate the relation between  the $A_0$ and $A_0^{cl}$ - the observable (physical) value of the $A_0$ condensate, which follows from calculation of $\left <  L \right> $. In course of these calculations we correct the results of Ref.\cite{bely91-254-153}, and find  the gauge-fixing independent EP $W_L(A_0^{cl})$, its minimum position and the value of PL in this vacuum. As application we calculate thermodynamical pressure and Debye's gluon mass. Discussions  and concluding remarks are given in the last section. In Appendix we carry out  the calculation of the one-loop correction to the PL omitted in Ref. \cite{bely91-254-153}, where only final result has been adduced. Also we present  information about Bernoulli's polynomials.

\section{Effective potential and Nielsen's identity}
Let us consider SU(2) gluodynamics in the Euclidean space time places  in the background field $\bar{A}^a_\mu = A_0 \delta_{\mu 0} \delta^{a 3} = const$ described by the Lagrangian
\be \label{Lagr} L = \frac{1}{4} (G^a_{\mu\nu})^2 + \frac{1}{2\xi}[(\bar{D}_\mu A_\mu)^a]^2 - \bar{C} \bar{D}_\mu D_\mu C. \ee
The gauge field potential $A_\mu^a = Q_\mu^a + \bar{A}^a_\mu $ is decomposed in quantum and classical parts. The covariant derivative in  Eq.\Ref{Lagr} is $(\bar{D}_\mu A_\mu)^{ab}= \partial_\mu\delta^{ab} - g \epsilon^{abc}\bar{A}^c_\mu,  G^a_{\mu\nu} = (\bar{D}_\mu Q_\nu)^a - ( \bar{D}_\nu Q_\mu)^a - g   \epsilon^{abc} Q^b_\mu  Q^c_\nu, g $   is a coupling constant, internal index a = 1,2,3. The Lagrangian of ghost fields $\bar{C}, C$  is determined by the background covariant derivative $  \bar{D}_\mu(\bar{A}) $ and the total one  $ D_\mu(\bar{A} + Q)$. As in Refs. \cite{bely90-45-355}, \cite{bely91-254-153} we introduce the ”charged basis” of fields:
\bea \label{fields}&& A^0_\mu =  A^3_\mu,~~ A^{\pm}_\mu = \frac{1}{\sqrt{2}} (A^1_\mu \pm i A^2_\mu), \\ \nn
&&  C^0 =  C^3,~~ C^{\pm} = \frac{1}{\sqrt{2}} (C^1  \pm i C^2 ). \eea
In this basis a scalar product is $x^a y^a = x^+ y^- + x^- y^+  + x^0 y^0$ , and the structure constants are:$\epsilon^{abc}$ = 1 for a = ”+”, b = ”-”, c = ”0”. Feynman's rules are the usual ones for the theory at finite temperature with modification: in the background field  a sum over frequencies should be replaced by $\sum_{k_0}, k_0 =( \frac{2 \pi n}{\beta} \pm g \bar{A}_0)$ in all loops of the fields $Q_\mu^{\pm}, C^{\pm}$ . Here,  $n = 0, \pm 1, \pm 2,...$ .  This frequency shift must be done not only in propagators but also in three particle vertexes.

Carrying out standard calculations we obtain the two-loop EP
\bea \label{EP} W(x) &=& W^{(1)}(x)+ W^{(2)}(x),\\ \nn
\beta^4 W^{(1)}(x)&=&  \frac{2}{3} \pi^2 [ B_4(0) + 2 B_4(\frac{x}{2}) ], \\ \nn
\beta^4 W^{(2)}(x)&=&\frac{1}{2} g^2 [  B_2^2(\frac{x}{2})+ 2 B_2(0)) B_2(\frac{x}{2})] +
\frac{2}{3} g^2(1 - \xi) B_3(\frac{x}{2})B_1(\frac{x}{2}),\eea
where $B_i(x)$ are Bernoulli's polynomials defined $ modulo$ 1  adduced in Appendix, $x = \frac{g A_0 \beta}{\pi}$. This expression coincides with calculated already in Refs.\cite{bely91-254-153}, \cite{skal92-7-2895}. In what follows we consider the interval $0 \leq x \leq 2$.

Let us investigate the minima of it. We apply an expansion in powers of $ g$ and get
\bea \label{W0} \beta^4 W_{min}& =&  \beta^4 W (0)
-\frac{1}{192\pi^2}(3-\xi)^2g^4, \nn \\
&&x =  g^2 \frac{(3 - \xi)}{8 \pi^2} ,\eea
where the first term is the value at zero field.
Actually, an expansion parameter determined from the ratio of two- and one-loop contributions equals to $\frac{g^2}{8 \pi^2}$, and therefore sufficiently large  coupling values are permissible.
As we see, both the minimum position and the minimum energy value are gauge-fixing dependent. Hence the gauge invariance of the $A_0$ condensation phenomenon is questionable.

This problem was solved within Nielsen's identity method in \cite{skal92-7-2895}, \cite{skal94-50-1150} for $SU(2)$ and $SU(3)$ gluodynamics and in \cite{skal94-9-4747}, \cite{skal94-57-324} for $ QCD$  with quarks.
Since this approach is important for what follows, we describe it in short here.

In Ref.\cite{kobe91-355-1} Nielsen's identity for general type EP has been derived:
\be \label{NI} \delta^{'} W (\phi) = W_{, i} \delta  \chi^i (\bar{\phi}), \ee
which describes a variation of $W(\phi)$ due to variation of the gauge fixing term $F^\alpha (\phi).$  In Eq.\Ref{NI} $\phi^i$ is gauge field, $\bar{\phi}^i$ denotes a vacuum value of $\phi^i$, comma after $W$ means variation derivative with respect to corresponding variable. Variation $\delta \chi^i  $ describes changing of field $(\bar{\phi})$  due to special gauge transformation which  compensates  variation of a classical action appearing after variation of gauge-fixing function $F^\alpha(\phi) \to F^\alpha(\phi) + \delta  F^\alpha(\phi)$.

In field theory $\delta \chi^i$ is caculated from equation \cite{kobe91-355-1}:
\be \label{varchi} \delta \chi^i = - \Bigl< D^i_\alpha(\phi) \Delta^\alpha_\beta(\phi)  \delta^{'}  F^\beta(\phi) \Bigr>, \ee
where $\bigl< O(\phi) \bigr> $ denotes functional average of $O(\phi)$. In this expression $D^i_\alpha(\phi)$ is generator of gauge group, $ \Delta^\alpha_\beta(\phi) $ is propagator of ghost fields, $ \delta^{'}  F^\beta(\phi)$ is variation of gauge fixing term.

In our case according Eq.\Ref{Lagr} $  \delta^{'}  F^\beta(\phi)  = - \frac{1}{2} (\bar{D}_\mu (\bar{ A}) Q_\mu)^\beta \frac{\delta \xi}{\xi}$,  $ D^i_\alpha$ is covariant derivative. In Ref.\cite{skal92-7-2895}, Eq. (26), the expression was derived (more details on calculations and discussions for $SU(3)$ case see in Refs.\cite{skal94-50-1150}, \cite{skal94-57-324}):
\be \label{varci} \delta \chi^0 = \frac{g}{4 \pi \beta} B_1(\frac{x}{2}) \delta \xi. \ee
Nielsen's identity  for two-loop EP reads
\be \label{NISU2} \frac{ d W}{d \xi} = \frac{\partial W^{(2)}}{\partial \xi} +  \frac{\partial W^{(1)}}{\partial x} \frac{\partial x}{\partial \xi} = 0, \ee
where in the order $\sim g^2$ the derivanive $\frac{\partial x}{\partial \xi}$ equals to $\frac{\delta \chi^0}{\delta \xi} \times ( \frac{g \beta}{\pi})$ in  Eq.\Ref{varci}. The latter factor comes from definition of $x = \frac{g A_0 \beta}{\pi}$.  Since $W^{(2)}$ has the order $g^2$, and  $W^{(1)}$-  $g^0$,
the Eq.\Ref{NISU2} states that $W(x, \xi)$ does not change along the charscteristic curve
\be \label{character} x = x' + \frac{g^2}{4 \pi^2} B_1(\frac{x'}{2}) (\xi - \zeta) \ee
in the plain of variables $(x, \xi)$, $\zeta $ is an arbitrary integration constant. Thus, there is the set of  orbits where $W(x')$ is gauge-fixing independent. Along them a variation in $\xi$ is compensated by the  special variation of $x'$.

\section{Effective potential of order parameter }
In this section we, following Ref.\cite{bely91-254-153}, express the EP \Ref{EP} in terms of  $\left< L \right> $. We call it "effective potential of  order parameter" $W_L(x_{cl})$.  In SU(2) group, in tree approximation, the PL is expressed in terms of $x$ as follows: $\left< L \right> $ = $\cos (\frac{\pi x}{2})$. This formula can be used to  relate physical value of  PL and  classical (observable) condensate value with accounting for radiation corrections: $\left< L \right> =\cos (\frac{\pi x_{cl}}{2}) =  \cos (\frac{\pi x}{2}) + \Delta\left< L \right> $. The quantum correction was calculated in one-loop order (Eq. (10) in Ref.\cite{bely91-254-153}),
\be \label{deltaPL} \Delta\left< L \right> = - \frac{g^2 \beta \sin (\frac{\pi x}{2})}{4 \pi} \int \frac{  d k  }{k_0^+} \bigl [ \frac{1}{( k_0^+)^2 + \vec{k}^2} + \frac{(\xi - 1)( k_0^+)^2 }{(( k_0^+)^2 + \vec{k}^2)^2 }\bigr], \ee
where the notations are introduced:
\be \int d k  = \int \frac{d^3 k}{(2 \pi )^3} \Bigl ( \frac{1}{\beta} \sum_{n = -\infty}^{\infty}  \Bigr ), ~~k_0^+ = k_0 + g A_0, ~~ k_0 = \frac{2 \pi n}{\beta}. \ee
Eq.\Ref{deltaPL}  is crucial for what folllows.

 In Ref.\cite{bely91-254-153} the  calculation of  $\Delta\left< L \right>$  was not  presented in detail and only the final expression for the EP  (Eq.(17))  has been given and analysed. We fill in this gap below.

   The second integral in Eq.\Ref{deltaPL} is well known, it is expressed in terms of Bernoulli's polinomials \cite{bely90-45-355}, \cite{skal94-57-324},
\be \label{I2} I_2 = - \frac{(\xi - 1)}{4 \pi \beta} B_1(\frac{x}{2}).\ee
Calculation of the first integral, $I_1$,  we reduce to the previous one.
Introducing the notation $\tilde{k}^2 =( k_0^+)^2 + \vec{k}^2$,  we write its integrand as follows
\bea \label{t1} Intgd. I_1& =&   \frac{  1  }{k_0^+}  \frac{1}{( k_0^+)^2 + \vec{k}^2} =   \frac{  k_0^+ }{[( k_0^+)^2 + \vec{k}^2]^2}  \frac{1}{1 - \vec{k}^2/\tilde{k}^2 } \nn\\
&=& \frac{  k_0^+ }{[( k_0^+)^2 + \vec{k}^2]^2} [ 1 + \sum_{l = 1}^{\infty} (\frac{\vec{k}^2}{\tilde{k}^2})^l ] . \eea
 Hence, the first term in   $I_1$ coincides (up to the factor$ (\xi - 1))$ with  $I_2$. The other terms are also positive. So, the sings of  $I_2$ and  $I_1$ must be the same.

 Now,  instead  summing   series over $l$  in Eq.\Ref{t1}, we return back to the initial expression for $ I_1$ and calculate it by using a standard procedure. This is presented in Appendix. The  result is
\be \label{I111}   I_1  = - \frac{1}{2 \pi \beta} B_1(\frac{x}{2}).\ee
Substituting  $I_1$ and $I_2$ in Eq. \Ref{deltaPL}, we obtain finally
\be \label{deltaPLa} \Delta\left< L \right> =  \frac{g^2  \sin (\frac{\pi x}{2})}{16 \pi^2}   B_1(\frac{x}{2} ) (\xi + 1) . \ee
Just this formula should be used in order to express    "nonphysical field" $ x$  in terms of "classical observable one", $x_{cl}$.

Note,  in  Ref.\cite{bely91-254-153}    the  final EP of order parameter  (Eq.(17))  corresponds to   the factor $(\xi - 3)$. But according to  Eq.\Ref{deltaPL}  the opposite  signs of $I_1$ and $I_2$  are impossible. Factor $( - 3 )$ can be obtained only for positive  sign in Eqs.\Ref{I111}, \Ref{I11}.

%
%


Actually, to get the  correct results in Ref. \cite{bely91-254-153}, we have to replace the parameter $(\xi - 3)$ by $(\xi + 1) $ in all the expressions. In particular, the relation between $ x$  and $x_{cl}$ looks as follows (compare with  Eq.(13) in Ref. \cite{bely91-254-153}):
\be \label{xclas} x = x_{cl} + \frac{g^2}{4 \pi^2} B_1 (\frac{x_{cl}}{2}) ( \xi + 1)  .\ee
 Within Nielsen's identity approach, this formula  corresponds to  the choice in Eq.\Ref{character}  $x' = x_{cl}$  and $\zeta =  - 1$.  Along this orbit the EP  is gauge-fixing independent and  expressed in terms of  $\left< L \right>$. In such a way these two methods are related.

 Inserting Eq.\Ref{xclas} in Eq.\Ref{EP} and expanding $B_4(\frac{x}{2})$ in powers of $g^2$,  we obtain $ W_L(x_{cl})= W^{(1)}_L(x_{cl})+ W^{(2)}_L(x_{cl}),$ where the first term is obtained from $W^{(1)}(x)$   by means of  substitution $ x \to x_{cl}$ and the second is
\be \label{EPL} \beta^4  W^{(2)}_L(x_{cl}) =  \frac{g^2}{2} \bigl [  B_2^2(\frac{x_{cl}}{2})+ 2 B_2(0) B_2(\frac{x_{cl}}{2}) +
 \frac{8}{3} B_3(\frac{x_{cl}}{2}) B_1(\frac{x_{cl}}{2})\bigr].\ee
In the $W_L(x_{cl})$ the  $\xi$-dependent terms are mutually cancelled, as it should be and demonstrate gauge-fixing independence.

 We also  note that the final  expression for $W_L(x_{cl})$    can be obtained from $W(x)$  Eq.\Ref{EP} formally (omiting described intermediate steps)  by means of the next substitutions: $x \to x_{cl}$ and $\xi \to \zeta = - 1$.

As a result, according Eq.\Ref{W0} we  get for the minimum values
\bea \label{Wf} \beta^4  W_L(x_{cl})|_{min}&=&  \beta^4 W_L (0)
-\frac{1}{48 \pi^2} g^4, \nn \\
x_{cl}|_{min}&=&  \frac{g^2}{2 \pi^2} .\eea
Thus, the EP  $W_L(x_{cl})$ has a nonzero minimum position  and does not depend on  $\xi$. The condensation happens at  the two-loop  level. The minimum  value of PL (corresponding to the physical  states)  equals to: $\left< L \right> = \cos( \frac{ g^2}{4 \pi}).$   In contrast,  in Ref.\cite{bely91-254-153} the value  $\left< L \right>  = \pm 1$ was obtained.

The expression $-  W_L(x_{cl})|_{min} = p$ \Ref{Wf}  gives a thermodynamical pressure in the plasma. The first  term is $  \beta^4 W_L (0) = - 0.657974 + \frac{g^2}{24}$.  The function $ W_L(x_{cl})$ can be used for calculating Debye's mass of neutral gluons defined as
\be \label{Dm} m^2_D = \frac{d^2 W_L(x_{cl}) }{d A_0^2}|_{A_0 = 0}, \ee
remind, $x_{cl} = \frac{g A_0^{cl}}{\pi T}.$ We get
\be \label{Dmgl} m^2_D = \frac{2}{3} g^2 T^2 + \frac{5}{4} \frac{g^4}{\pi^2}  T^2. \ee
Here, first term is well known one-loop contribution and the second is two-loop correction.

To complete we note that the $A_0$ condensation is derived within the  correlation of the one- and two-loop  effective potentials. Whereas asymptotic freedom at high temperature is realized due to the  relation of the tree-level  and one-loop contributions to the EP.  Formally, the latter  results in the replacement of coupling constant  $g^2 \to \bar{g}^2 \sim \frac{g^2 }{\log(T/T_0)}$, $T_0$ is a reference temperature.  In both cases, the ratio of the relevant terms  is $\sim \frac{g^2}{4 \pi^2}$.  Hence it follows that at high temperature  we can substitute $g^2 \to \bar{g}^2$ in all the above formulas, in particular, in Eq. \Ref{Wf}.

Thus, the value of the order parameter PL in the minimum of the EP is
\be \label{PLg} \left< L \right> = \cos( \frac{ \bar{g}^2}{4 \pi}).\ee
It gives a possibility for determination of the deconfinement phase transition and its type. Accounting for the explicit expression for the one-loop effective coupling $\alpha_s = \frac{ \bar{g}^2}{4 \pi}$ in the $SU(2)$ case
\be \label{alpha} \bar{\alpha}_s = \frac{\alpha_s}{1 + \frac{11}{3 \pi} \alpha_s \log(T/T_0)}\ee
we see that the PL is continuously decreasing with  temperature lowering and becomes zero at $ \frac{ \bar{g}^2}{4 \pi} = \frac{\pi}{2}$. This signals a confinement. If we set $T_d = T_0$ the value of the ratio $W^{(2)}/W^{(1)} $ is $\sim 1/2$, that is in the range of applicability of perturbation theory. The phase transition is second order, as it is well known for this gauge group. We stress once again that due to the smallness of the expansion parameter our perturbation  effective potential of order parameter is suitable for investigating the confinement phase transition.
\section{$A_0$ condensate and stabilization of magnetic fields }
During recent years  it was realized that in plasma strong temperature dependent chromomagnetic  fields have to be spontaneously generated \cite{skal96-387-835}, \cite{star94-322-403}, \cite{Demc15-46-1}. These
phenomena  are   related   with  asymptotic freedom  in covariantly constant  fields  and take place even at zero temperature \cite{savv77-71-133}. They are realized   because  for  such  fields  an  infrared region of   strengths  has to be  unstable.  In fact, this is   Savvidy's vacuum at finite temperature. In contrast to the zero temperature case, at finite temperature magnetic field stabilization takes place. Moreover, the $A_0$ classical fields   act as stabilizing factors \cite{star94-322-403}, \cite{demc08-41-164051}.   As a result, the  background of plasma is not a perturbative  vacuum.  The actual one is formed out of    gauge field condensates. Moreover, the presence of magnetic fields is an other independent  signal of the phase transition.  In general, we have to take into consideration both type of condensates.

At  high temperature, when perturbation methods are reliable,  it is possible to analyze the role of the $A_0$ condensate as infrared stabilizing factor. Note that this idea first was proposed in Ref. \cite{star94-322-403}. Now, when the value of the condensate is derived \Ref{Wf}, we can demonstrate that using the result of Ref.\cite{skal00-576-430} (Eq.(19)), where the spontaneously created chromomagnetic field was calculated  for $SU(2)$ gluodynamics:
\be \label{H} (g H)_{cl} = \frac{g^4}{4 \pi^2} T^2.\ee
On the other hand, the value of the $A_0$ condensate following from Eq. \Ref{Wf} is
\be \label{A0cl} (A_0)_{cl}|_{min} = \frac{g^2}{2 \pi} T.\ee
Hence, for the charged gluon spectrum in the classical background fields directed along third axis in internal and usual spaces,  we get
\be \label{Spectrum} (p_4 + \frac{g^2}{2 \pi} T)^2 + p^2_{3} + (2 n - 1)\frac{g^4}{4 \pi^2} T^2, \ee
where $p_{3}$ is momentum along the field $H(T)$  and $n = 0,1,2,..$ is Landau level number. In the ground state $n = 0$ and the T-depended term in the first brackets cancels the negative last one. The spectrum is stable. In contrast, for $A_0 = 0$ the well known unstable mode $ p_4^2 + p^2_{3} - g H $ is reproduced. $A_0$ condensate stabilizes infrared region of momenta $p_3 \to 0$. It worth to note that in Ref.\cite{demc08-41-164051} the stabilization was detected in simulations on the lattice for some chosen values of the $A_0 \not = 0$ condensate.  Here we presented it for the values of fields obtained in the consistent analytic gauge invariant calculations. In Ref.\cite{demc08-41-164051} it was also found that at low temperature the magnetic field is absent (screened).

 \section{Discussion and conclusions}
Two main conclusions follow from the above considerations. First, we found a simple correlation between Nielsen's identity method  for the effective potential \Ref{EP} and the EP of order parameter $W_L(x_{cl})$. To link them we have to substitute in Eq.\Ref{EP} $ x \to x_{cl}, \xi \to \zeta = - 1$.  We see that the EP satisfying Nielsen's identity is $\xi$-independent on numerous orbits in the $(x, \xi)$-plain.
 The only free parameter  is the value $\zeta$ in Eq.\Ref{character}.  The found choice,  which  relates these approaches  for describing gauge invariance, is $\zeta = - 1$. Second, we calculated  the observable "classical" values of the condensate  $ x_{cl}|_{min}$ Eq.\Ref{Wf}. In fact, we just  realized the  Belyaev's idea  of expressing the EP in terns of  $\left< L \right>$.

It worth to say a few words about the orbit with $\zeta  = 3$  in Eq.\Ref{character},  that corresponds to the case considered in Ref.\cite{bely91-254-153}. This is a special case where $x_{cl} = 0$ at two-loop order.  Actually, zero  value here separately  follows  either for  the one-loop EP or the two-loop  one plus the terms $\sim g^2 $ coming from the expansion of $B_4(\frac{x}{2})$ in the former EP. 	Looking at Eq.\Ref{W0},  in  the  gauge  $\xi = 3$ we have to expect  $A_0$ condensation in three-loop approximation. But in general basically  this  is two-loop phenomenon.  The  standard loop expansion method is well applicable for this problem.

As we noted in Introduction, in Ref.\cite{kort20-803-135336} the $\xi$-independence of the $A_0$ has been established within the constrained EP approach. It  is close to the  EP of order parameter and results in the same conclusions. The main difference between them consists in the calculation procedures applied. In the former one the PL is considered as a dynamical parameter. In the latter the dynamical parameter is $A_0$ external field which is a solution to local field equations.  This field  is expressed though  $\left< L \right>$ after actual calculation of the EP $W(A_0)$. It is also important that qualitatively these results are in agreement with the ones derived by means of lattice QCD methods \cite{bori91-264-166}, \cite{Bori97-56-5086}. In these papers the EP for the  PL has been studied in a gauge invariant lattice formulation.

In is also interesting to compare our results with the ones in Ref.\cite{rein15-742-61} where the DPT in $SU(2)$ gluodynamics was investigated by a perturbation method. Special modified Landau-DeWitt gauge  was applied. It differs from the original  one by an addition mass term $L_m$, introduced in a standard way
\be \label{Lm} L_m = \frac{1}{2} m^2 Q^a_\mu Q^a_\mu.\ee
This term is considered by the authors as an infrared cut-off for gluon low momenta. The quantization has been carried out by means of the background field method. In fact, such a simple introducing of mass term for Yang-Mills fields has to result in a nonrenormalizable theory, which is nonunitary. This will be discussed in details in other place.
Here in short we mention that the only way to preserve renormalizability is spontaneous symmetry breaking when, in particular, the contribution of the scalar components  of massive vector fields (having indefinite metrics)   is cancelled by the contributions of scalar fields introduced via the Higgs-Kibble mechanism. Without such cancellations the theory should be cut-off at large momenta. The limit $m \to 0$ is singular. Other remark  is that the introduction of the mass term  suppresses the solution  $H = const$ to field equations - Savvidy's vacuum, which is proper to massless case, only. The mass parameter $m$ screens  classical long range  field, and no  magnetic background could be detected. As a result, important dynamical phenomenon regulating low momenta dynamics of Yang-Mills theory is ruled out in this gauge.

 Now, let us compare the results relevant to our present investigation.
In Refs. \cite{rein15-742-61}, \cite{rein16-93-105002} the applicability of perturbation methods in deconfinement phase transition is also motivated by the smallness of the expansion parameter $\frac{g^2}{4 \pi^2}$ in the region of the transition. The presence of the $A_0$ condensate in the high temperature phase was also observed. In contrast to our investigation the low temperature phase has been taken into consideration, that is impossible in our case and our results are reliable till the temperatures whet PL becomes zero. Remind that this happens when $g^2 = 2 \pi^2$ and the expansion parameter  equals  $\frac{1}{2}$. Qualitatively they are in agreement with each other.

It is also very important that the presence of the $A_0$ condensate stabilizes temperature dependent magnetic fields. Hence it follows that the stable  background of the plasma is formed out of two classical condensates $A_0$ and $H(T)$. These fields occupy the whole  space volume of the plasma. The  magnetic mass of the background magnetic field is zero. And electric Debye's mass of neutral gluons is finite and given by Eq.\Ref{Dmgl}.

Summarizing all together, we have proven   explicitly the   gauge-fixing independence of the $A_0$ condensate. This condensate acts az the dynamical parameter regulating behavior  of gauge fields in the  infrared region of momenta.
The simplicity of dealing with constant background potentials $A_0$ makes them  useful and convenient objects for different applications in QCD after deconfinement phase transition.

The author grateful Michael  Bordag and Oleg Borisenko for constructive remarks and suggestions.

\section*{Appendix}

Let us calculate $I_1$.
For the temperature sum we use the representation
\be \label{tsum} S^{(0)} =  \frac{1}{\beta} \sum_{n = -\infty}^{\infty} F(k_0) = \frac{1}{4 \pi i} \int_C \cot (\frac{1}{2} \beta \omega)  F(\omega) d \omega - \frac{1}{2 \pi} \int_{  -\infty}^{\infty} F(\omega) d \omega, \ee
where  $k_9 = \frac{2 \pi n}{\beta} $  and contour $C$ goes opposite clock-wise around the real axis in complex $\omega$-plane where the poles at $k_0$ are located. The second integral removes  a zero temperature contribution. To calculate the contour integral we have to extend $C$ to infinity and calculate the sum of residuums,
\be \label{I0} I^{(0)} = - \frac{1}{2} \sum Res[ \cot \bigl(\frac{1}{2} \beta \omega\bigr)  F(\omega)]. \ee
The sign "-" reflects that the poles of $F(\omega)$ are passed round clock-wise in the $\omega$-plane.

In our case (see Eq.\Ref{deltaPL}), $F(\omega)$ has two simple poles: $\omega = - g A_0 \pm i k$, and we get
\bea \label{I01} I_1^{(0)} &=& -\frac{1}{4 k^2} \bigl( \cot [\frac{1}{2} \beta (g A_0 + i k) ] -  \cot [\frac{1}{2} \beta ( - g A_0 + i k)] \bigr) \nn \\
&=& \frac{i}{4 k^2} \Bigl[ \frac{1 + e^{i\beta (g A_0 + i k )}}{1 - e^{i\beta (g A_0 + i k )}} -  \frac{1 + e^{i\beta ( - g A_0 + i k )}}{1 - e^{i\beta (-g A_0 + i k )}} \Bigr].\eea
Introducing the notations $ X_1 = e^{- \beta k + i \beta g A_0 },  X_2 = e^{- \beta k - i \beta g A_0}$, the  Eq.\Ref{I01} can be written in the form
\be \label{I02}   I_1^{(0)} = \frac{i}{2 k^2} \sum_{ n =1}^{\infty} ( X_1^n -  X_2^n) = - \frac{1}{ k^2} \sum_{ n =1}^{\infty} e^{- \beta k n} \sin( g A_0 \beta n). \ee
Then, performing momentum integration and summing up the series, we obtain
\be \label{I11}   I_1  = \frac{1}{2 \pi^2} \sum_{ n =1}^{\infty} \frac{1}{\beta n} \sin( g A_0 \beta n) = - \frac{1}{2 \pi \beta} B_1(\frac{x}{2}),\ee
where $x = \frac{g A_0 \beta}{\pi}$.

The Bernoulli's polynomials defined $ modulo$ 2   are
\bea \label{BP} B_1(x)&=&x - \frac{x}{2 |x|}, ~~ B_2(x) = x^2 - |x| + \frac{1}{6}, \\ \nn
B_3(x)&=& x^3 - \frac{3}{2} \frac{x^3}{|x|} + \frac{1}{2} x, \\ \nn
B_4(x)&=& x^4 - 2 |x|^3 + x^2 - \frac{1}{30}. \eea
At $x = 0 $ the $B_1(x)$ is defined to be 0.


\end{document}